\documentclass{svjour3}                     
\smartqed  
\usepackage{graphicx}
\usepackage{epstopdf}
\usepackage{rotating}
\usepackage{mathptmx}      
\usepackage{natbib}
\journalname{SSRv}

\newcommand{\mum}{$\mathrm { \mu m }$}
\def\la{~\raise.4ex\hbox{$<$}\llap{\lower.6ex\hbox{$\sim$}}~} 
\def\ga{~\raise.4ex\hbox{$>$}\llap{\lower.6ex\hbox{$\sim$}}~}
\def\Rgd{\hbox{$\mathrm{R_{g/d}}$}}
\def\NHI{\hbox{N(H$^{\rm o}$)}}
\def\NH2{\hbox{N(H$_{\rm 2}$)}}
\def\FeII{\hbox{Fe$^{\rm +}$}}
\def\SII{\hbox{S$^{\rm +}$}}
\def\SiII{\hbox{Si$^{\rm +}$}}
\begin{document}

\title{Interstellar Dust Inside and Outside the Heliosphere
}

\titlerunning{Interstellar Dust}        

\author{Harald Kr\"uger          \and
        Eberhard Gr\"un
}

\authorrunning{H. Kr\"uger and E. Gr\"un} 

\institute{H. Kr\"uger \at
 Max-Planck-Institut f\"ur Sonnensystemforschung,
 Max-Planck-Str. 2,  37191 Katlenburg-Lindau, 
 Germany, Tel.: +49-5556-979234,
              Fax:  +49-5556-979240,
              \email{krueger@linmpi.mpg.de}           
           \and
           H. Kr\"uger \and E. Gr\"un \at
              Max-Planck-Institut f\"ur Kernphysik,
              Saupfercheckweg 1,
              69117 Heidelberg, 
              Germany 
              \and 
              E. Gr\"un \at Laboratory for Atmospheric and Space Physics, University of Colorado, 
Boulder, CO, 80303-7814, USA
              }

\date{Received: date / Accepted: date}

\maketitle

\begin{abstract}
In the early 1990s, after its Jupiter flyby, 
the Ulysses spacecraft identified interstellar dust in
the solar system. Since then the in-situ dust detector
on board Ulysses continuously monitored interstellar grains 
with masses up
to $\rm 10^{-13}$\,kg, penetrating deep into the solar system.
While Ulysses measured the interstellar dust stream at high ecliptic 
latitudes between 3 and 5~AU, interstellar impactors were also measured 
with the in-situ dust detectors on board Cassini, Galileo and Helios, 
covering a heliocentric distance range between 0.3 and 3 AU in the 
ecliptic plane.
The interstellar dust stream in the inner solar system
is altered by the solar radiation pressure force, gravitational focussing and
interaction of charged grains with the time varying interplanetary 
magnetic field.
The grains act as tracers of the physical conditions 
in the local interstellar cloud (LIC).
Our in-situ measurements imply the existence of a population 
of 'big' interstellar grains (up to $\mathrm{10^{-13}\,kg}$) and 
a gas-to-dust-mass ratio
in the LIC which is a factor 
of $>$ 2 larger than the one derived from astronomical observations, 
indicating a concentration of interstellar dust in the very 
local interstellar medium. 
Until 2004, the interstellar dust 
flow direction measured by Ulysses was
close to the mean apex of the Sun's motion through
the LIC, while   
in 2005, the data showed a $30^{\circ}$ shift,  
the reason of which is presently unknown. We review the results from 
spacecraft-based in-situ 
interstellar dust measurements in the solar system
and their implications for the physical and chemical state of the LIC.

\keywords{dust \and interstellar dust \and heliosphere \and interstellar matter}

\end{abstract}

\section{Introduction}
\label{intro}
Interstellar dust (ISD) became a topic of astrophysical research
in the early 1930s when the existence of extinction, weakening, and scattering
of starlight in the interstellar medium (ISM) was realised. At that time, astronomical 
observations provided the only information about the properties of the dust in the ISM.
With the advent of dust detectors onboard spacecraft, it became possible to 
investigate dust particles in-situ. About 30 years ago, analysis of the data
obtained with the dust instruments flown on a couple of spacecraft suggested 
that ISD grains can cross the heliospheric boundary and penetrate deeply 
into the heliosphere \citep{bertaux1976,wolf1976}. In the 1990s, this 
was undoubtedly demonstrated with the dust instrument carried by the Ulysses
spacecraft: the Ulysses dust detector, which measures mass, speed and approach 
direction of the impacting grains, identified ISD grains with radius
above 0.1\,\mum\, sweeping through the heliosphere 
\citep{gruen1993a,gruen1994a,gruen1995a}. 

\begin{figure}
\begin{center}
\includegraphics[scale=0.75]{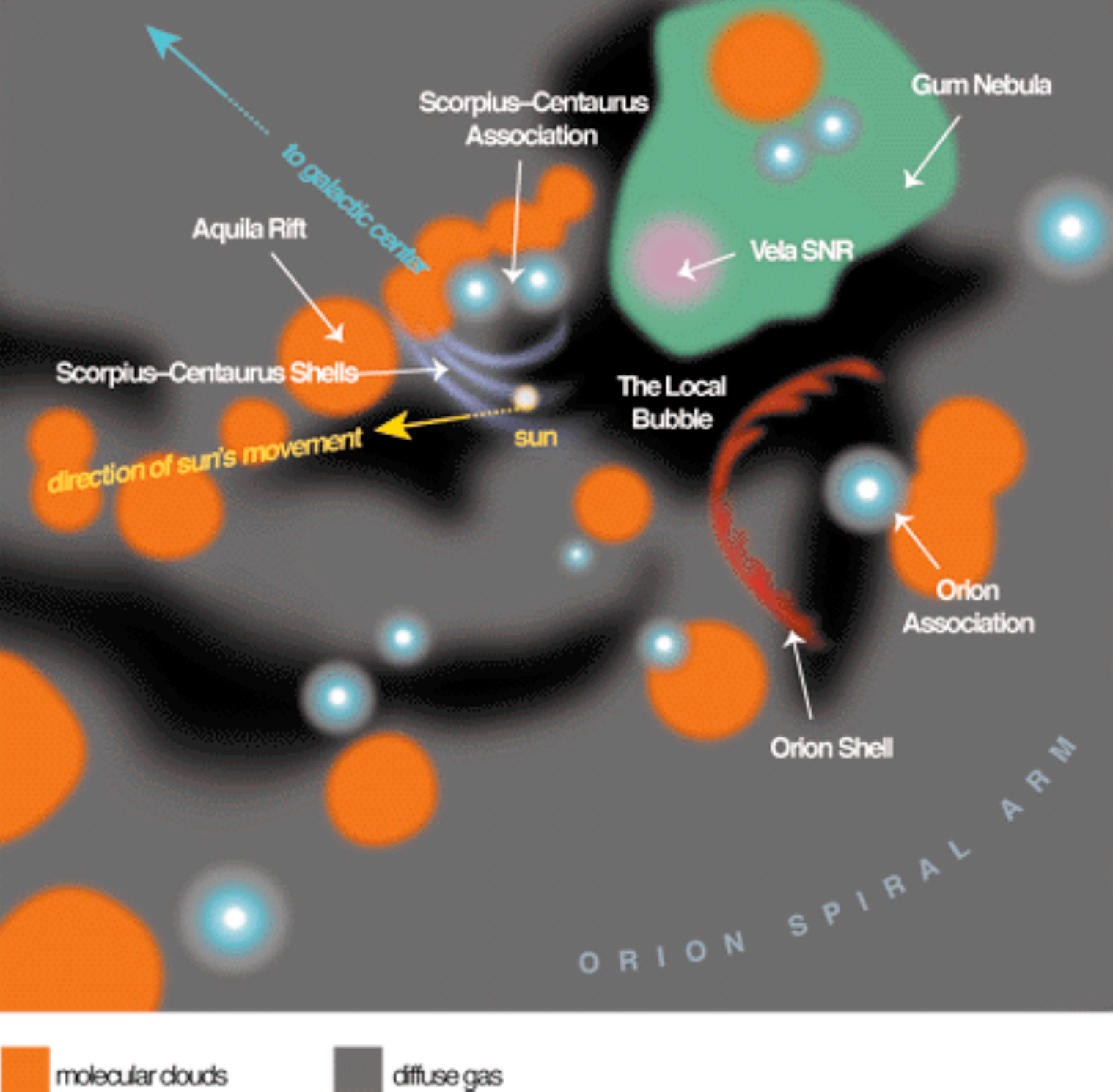}
\end{center}
\caption{\label{Frisch}
Our galactic environment within 500~pc of the Sun. 
Currently, the Sun is passing through the Local Interstellar Cloud (LIC), 
shown in violet, which is flowing away from the Scorpius-Centaurus Association 
of young stars. The LIC resides in a low-density hole in the interstellar medium 
called the Local Bubble, shown in black. 
Nearby, high-density molecular clouds including the Aquila Rift surround star forming 
regions, each shown in orange. The Gum Nebula, shown in green, is a region of hot 
ionized hydrogen gas. Inside the Gum Nebula is the Vela Supernova Remnant, shown in 
pink, which is expanding to create fragmented shells of material like the LIC 
(from P.\,C. Frisch, University of Chicago).
}
\end{figure}

The galactic environment of our solar system on a larger scale is shown in
Figure~\ref{Frisch}. 
The nearby interstellar medium (within about 3~pc of the Sun) is dominated by a
shell of material, the Local Interstellar Cloud (LIC). The solar system 
currently passes through the LIC which
is located at the edge of the Local Bubble.  The Local Bubble was
excavated by supernova explosions in the neighbouring
starforming regions of the Scorpius-Centaurus and Orion Associations. 
The solar system emerged 
from the interior of this bubble within the last $\mathrm{10^5}$ years. 
The only direct observation of ISD close to the Sun is weak polarization
observed along the sightline towards 36 Oph (distance about 6~pc) which is due
to magnetically aligned dust grains \citep{tinbergen1982}. Therefore, 
in-situ sampling of dust from the LIC can
greatly improve our understanding of the nature and processing of dust in 
various galactic environments and can cast new light on the chemical composition
and homogeneity of the interstellar medium. 

In this paper we review the results from in-situ ISD measurements
obtained with the Ulysses and other space-borne dust detectors. We review our 
current knowledge about ISD inside the heliosphere and in our local
interstellar environment. 

\section{Interstellar dust inside the heliosphere}

The Ulysses in-situ dust measurements showed that the grain motion through 
the solar 
system is parallel to the flow of neutral interstellar hydrogen and helium gas, 
both gas and dust travelling with a speed of $\mathrm{ 26\,km\,s^{-1}}$
\citep{gruen1994a,baguhl1995a,witte1996,witte2004}. 
The upstream direction of the dust flow lies at $\mathrm{259^{\circ}}$ ecliptic
longitude and $\mathrm{8^{\circ}}$ latitude \citep{landgraf1998a}.
The interstellar dust flow persists at high ecliptic latitudes above and below
the ecliptic plane and even over the poles of the Sun, whereas 
interplanetary dust is strongly depleted at high latitudes 
\citep{gruen1997a}.  The interstellar dust flux measured at a distance of 
about 3~AU from the Sun is time-dependent, and the mean mass of the grains
is about $\mathrm{3\cdot 10^{-16}\,kg}$ \citep{landgraf2000a}, corresponding
to a grain radius of approximately 0.3 \mum. 
Measurements with the identical dust instrument onboard 
Galileo performed in the ecliptic plane showed that beyond about 
3~AU the interstellar dust flux even exceeds 
the flux of micron-sized interplanetary grains.

Results from the first decade of the Ulysses mission showed that
the radii of clearly identified interstellar grains range from
0.05\mum\, to above 1\mum. 
The data show distance-dependent 
alteration of the interstellar dust stream caused by gravitational focussing by the Sun, 
solar radiation pressure, and electromagnetic interaction 
with the time-varying interplanetary magnetic field 
\citep[IMF;][]{altobelli2003,altobelli2005b,altobelli2005a,mann2000a,
landgraf2000b,czechowski2003}.
Radiation pressure and electromagnetic forces strongly depend on grain size,
leading to a strong modification of the size distribution and fluxes of 
grains measured inside the heliosphere \citep{landgraf1999a,landgraf2003}.
The size distribution 
shows a deficiency of small grains below 0.3\mum\, compared to astronomically
observed ISD \citep{frisch1999a}. 
In addition, solar radiation pressure deflects grains with
sizes of about 0.4\mum\, and was found to be effective at solar distances below
4~AU \citep{landgraf1999a}.

Significant differences in the particle sizes were also recorded at different 
heliocentric distances. In addition to the Ulysses measurements which revealed 
a lack of small
0.3\mum\, ISD grains within 3~AU heliocentric distance, measurements by
Cassini and Galileo in the distance range between 0.7 and 3~AU showed that 
the detected interstellar particles were bigger than 0.5\mum, with grain 
masses increasing closer to the Sun \citep{altobelli2003,altobelli2004b,altobelli2005a}. 
The flux of these bigger particles did not
exhibit temporal variations due to the solar-wind magnetic field like the flux
of smaller particles observed by Ulysses. The trend of increasing 
particle masses continues even closer to the Sun, as demonstrated by Helios
which recorded particles of about 1\mum\, down to 0.3~AU \citep{altobelli2005b,altobelli2006}.
These facts support the idea that the ISD stream is strongly filtered by solar 
radiation pressure. Interstellar particles with optical properties 
of astronomical silicates or organic refractory materials are
consistent with the observed radiation pressure effects \citep{landgraf1999a}.

\begin{figure}[t]
\begin{center}
\vspace{-16.5cm}
\parbox{0.99\hsize}
{\hspace{-0.3cm}
\includegraphics[scale=0.9]{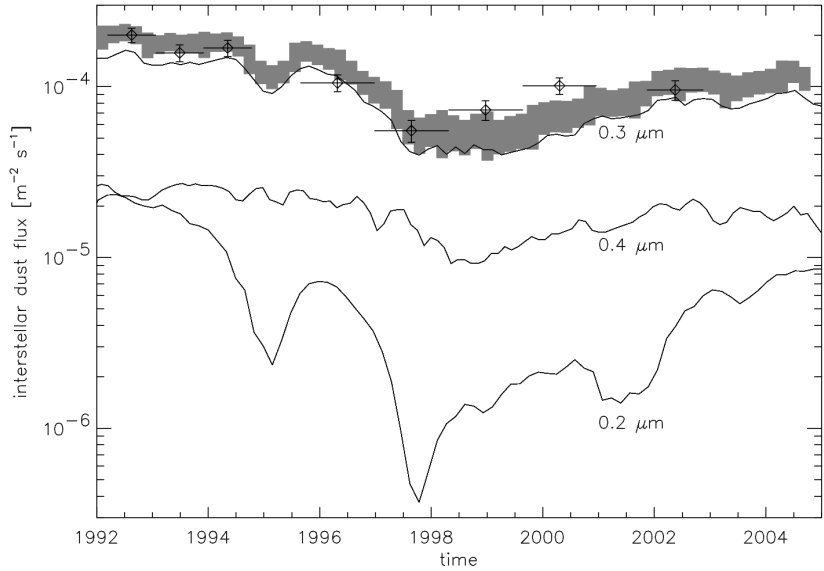}
\vspace{-0.5cm}
}
\end{center}
\caption{
Fit of simulated to measured flux \citep{landgraf2003}. The fit
parameters are the relative
contributions of grains of sizes between 0.1 and 0.4 $\mu$m (the
0.1~$\mu$m curve
is not shown, because it did not contribute to the fit). The solid lines
show
the flux profiles of the simulated grains of various sizes, scaled with
their
best-fit relative contributions. The shaded region indicates the
best-fit total predicted
flux, with its vertical extent giving the $1\sigma$
uncertainty.\label{fig_fit}
}
\end{figure}

In addition to studies of the distribution of grain masses, the Ulysses dust
instrument monitors the flux of the interstellar particles in the heliosphere 
(Figure~\ref{fig_fit}). In mid 1996, we observed a decrease of the
interstellar dust flux by a factor of 3 from an initial value of $\mathrm{1.5\times 10^{-4}\:{
m}^{-2}\:{s}^{-1}}$ down to $\mathrm{0.5\times 10^{-4}\:m^{-2}\:s^{-1}}$.
This drop was attributed to increased
filtering of small grains by the solar wind driven IMF 
during solar minimum conditions \citep{landgraf1998a,landgraf2000b,landgraf2000a}. 
Since early 2000, Ulysses has again detected
interstellar dust flux levels above $\mathrm{10^{-4}\:m^{-2}\:s^{-1}}$
\citep{landgraf2003,krueger2007b}.
Monte-Carlo simulations of the grain dynamics in the heliosphere 
showed that the dominant contribution to the dust flux
comes from grains with a charge-to-mass ratio of 
$\mathrm{q/m=0.59\:C\:kg^{-1}}$ and a
radiation pressure efficiency of $\mathrm{\beta=1.1}$,  corresponding to
grain radii of $\mathrm{0.3\:\mu m}$ \citep{landgraf2003}.

Particles even bigger (40\mum) than the grains measured in-situ with the 
spacecraft detectors were reliably identified by meteor radar observations 
\citep{taylor1996b,baggaley2000,baggaley2002,meisel2002}. The grains were identified by their
hyperbolic speeds, and their flow direction varies over a much wider angular range 
than that
of the much smaller grains observed by spacecraft. \citet{baggaley2000} identified a
general background influx of extra-solar system particles from southern ecliptic 
latitudes with enhanced fluxes from discrete sources. More sensitive meteor observations 
with the Arecibo radar found micron-sized interstellar meteor particles 
radiating from the direction of the Geminga pulsar \citep{meisel2002}. This is 
particularly interesting because the supernova
that formed the Geminga pulsar is a potential candidate which may have created
the Local Bubble. 

\begin{figure}[t]
\begin{center}
\vspace{-0.5cm}
\hspace{-1.2cm}
\parbox{0.49\hsize}
{
\begin{turn}{180}
\includegraphics[scale=0.25]{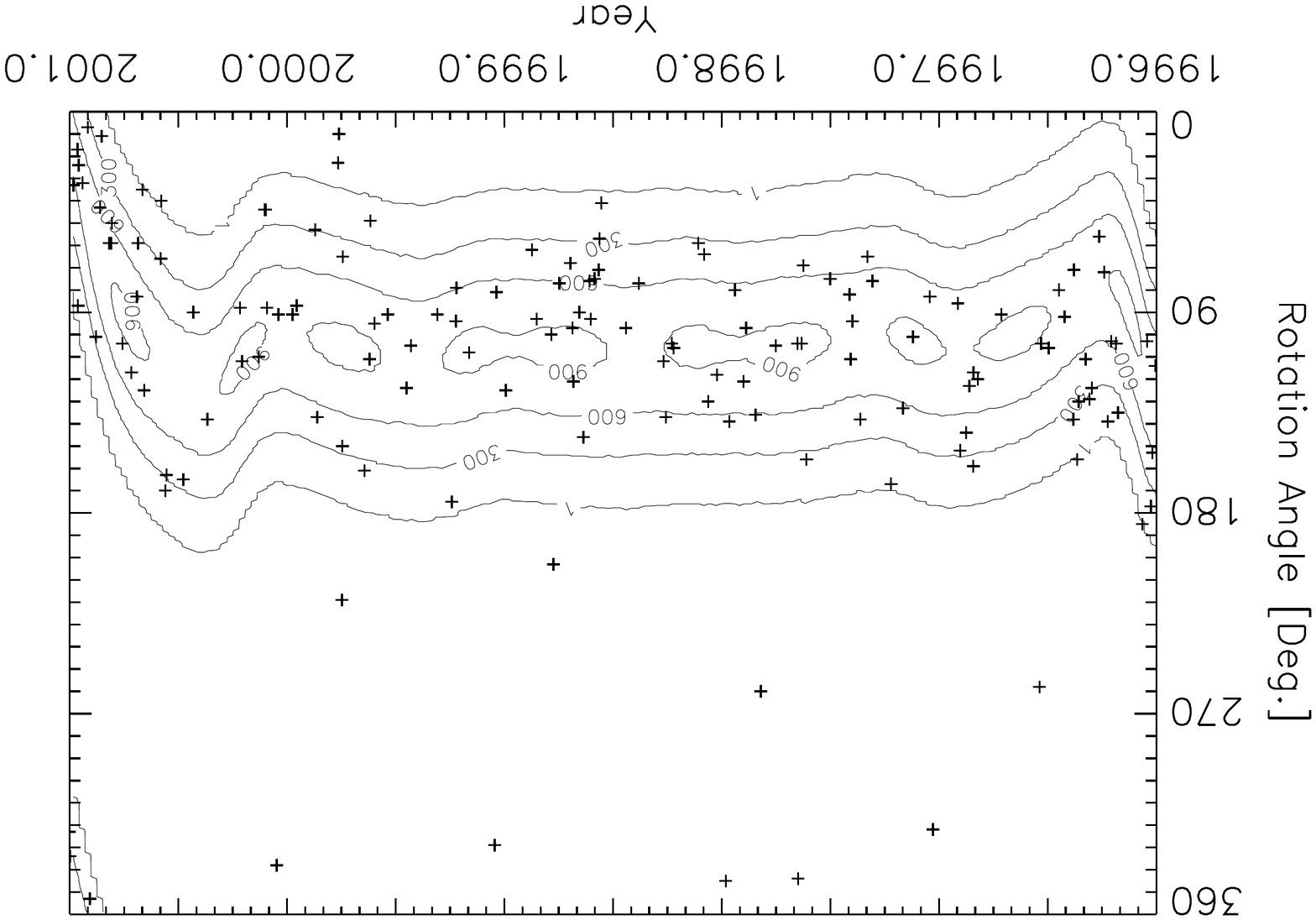}
\end{turn}
}
\parbox{0.49\hsize}
{
\begin{turn}{180}
\includegraphics[scale=0.25]{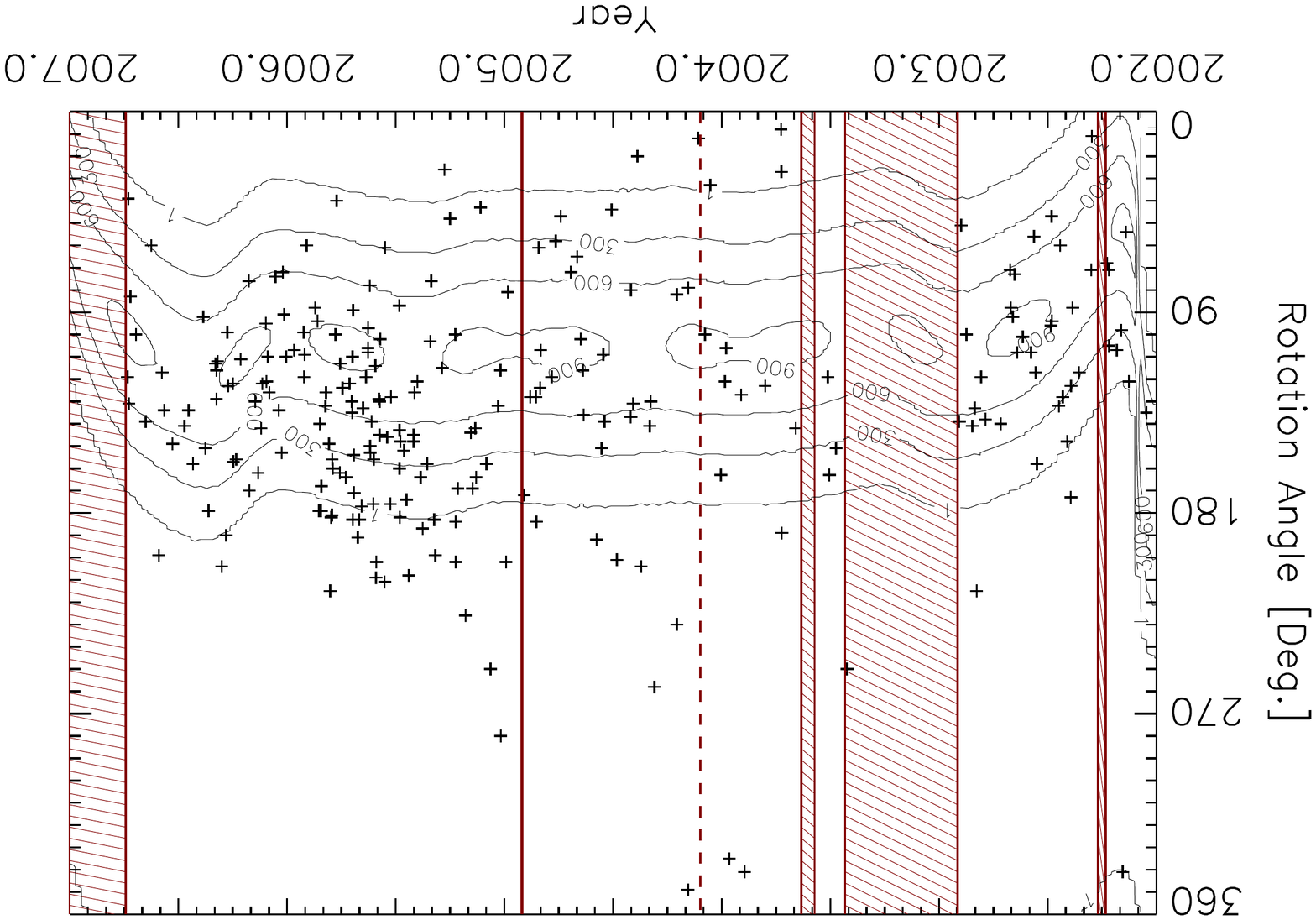}
\end{turn}
}
\vspace{-0.4cm}
\caption{\label{rot}
Impact direction (i.e. spacecraft rotation angle at dust particle impact) 
of interstellar grains measured with Ulysses in two time intervals 
\citep[from][]{krueger2007b}. {\em Left:} 1 January 1996 to 31 December 2000;
{\em right:} 1 January 2002 to 31 December 2006.
Ecliptic north is close to $\mathrm{0^{\circ}}$. 
Each cross indicates an individual impact.
Contour lines show the effective sensor area for particles
approaching from the upstream direction of interstellar 
helium. In the right panel, a vertical dashed line shows Jupiter 
closest approach on 5 February 2004, five shaded areas indicate 
periods when the dust instrument was switched off.
}
\end{center}
\end{figure}

Ulysses has monitored the interstellar dust flow through the 
solar system for more than 15 years now. This time period covers 
more than two and a half revolutions of the spacecraft about the
Sun through more than 2/3 of a complete 22-year solar cycle. Thus,
Ulysses measured interstellar dust during solar minimum and 
solar maximum conditions of the interplanetary magnetic field (IMF).
The interstellar dust flux modulation due to grain interaction with 
the rather undisturbed magnetic field during
solar minimum could be well explained 
\citep{landgraf1998a,landgraf2000b,landgraf2003}. The consideration of
sensor side wall effects lead to an improved flux determination 
\citep{altobelli2004a}. 

Until early 2005 the approach direction of the interstellar 
grains was
in agreement with the interstellar helium flow direction 
\citep{landgraf1998b,frisch1999a,krueger2006b}.
An example is shown in the left panel of Figure~\ref{rot} which
shows the impact direction of the interstellar impactors in the
period from 1996 to 2000. Six years later, when Ulysses 
was travelling through almost the same spatial region and had
an almost identical detection geometry for interstellar grains, the situation
was vastly different: first, the range in approach directions of the 
grains was somewhat wider (best seen in 2004; right panel of Figure~\ref{rot}; see 
also \citet{krueger2007b}); 
second, and more noticeable, in 2005/06 the approach direction of the majority of grains
was shifted away from the helium flow direction. Preliminary analysis 
indicates that this shift is about $\mathrm{30^{\circ}}$ away from 
the ecliptic plane towards southern ecliptic latitudes \citep{krueger2007b}.
At the moment, we do not know whether
it is a temporary shift limited to the time period stated above 
or whether it continues to the present time. Furthermore, 
the reason for this shift remains mysterious. Whether it is connected
to a secondary stream of interstellar neutral atoms shifted from
the main neutral gas flow \citep{collier2004,wurz2004,nakagawa2006} is presently 
unclear.
Given, however, that the neutral gas stream is shifted along the ecliptic
plane while the shift in the dust flow is offset from the ecliptic,
a connection between both phenomena seems unlikely.

Even though Ulysses' position in the heliosphere and the dust detection conditions were very similar during both time intervals considered in Figure~\ref{rot}, the configurations of the solar wind driven interplanetary magnetic field (IMF), which strongly affects the dynamics of the smallest grains, were completely different. We have to consider that the interstellar grains need approximately twenty years to travel from the heliospheric boundary to the inner solar system where they are detected by Ulysses. Thus, the effect of the IMF on the grain dynamics is the accumulated effect caused by the interaction with the IMF over several years: In the earlier time interval (1997-1999) the grains had a recent dynamic history dominated by solar minimum conditions \citep{landgraf2000b}, while the grains detected during the second interval (2002-2005) had a recent history dominated by the much more disturbed solar maximum conditions of the IMF. During the solar maximum conditions the overall magnetic dipole field changed polarity. \citet{morfill1979b}  predicted that due to this effect in a 22-year cycle, small interstellar grains experience either focussing or defocusing conditions. During these times they are systematically deflected by the solar wind magnetic field either towards or away from the solar magnetic equator plane (close to the ecliptic plane). This latter configuration likely has a strong influence on the dust dynamics and the total interstellar flux in the inner heliosphere but it is not modelled in detail in the presently existing models. An explanation of the grain interaction with the IMF at the recent solar maximum conditions is still pending.

The fact that the models fit the flux variation by assuming 
a constant dust concentration in the Local Interstellar Cloud (LIC) 
implies that the dust phase of the LIC must be homogeneously distributed 
over length scales of $\mathrm{50\:AU}$, which is the distance inside 
the LIC travelled by the Sun during the measurement period of 
Ulysses from end 1992 to end 2002 \citep{landgraf2003}. This conclusion
is supported by the more recent Ulysses data until the end of 2004 
\citep{krueger2006b}. The 2005/06 data, on the other hand, put a question 
mark onto this 
conclusion because if the observed shift in impact direction turns
out to be intrinsic, it would imply that this homogeneity breaks down on 
larger length scales.

\section{Interstellar dust in the Local Interstellar Cloud}

ISD grains carry 
information about their past dynamics outside the heliosphere and are thus of 
strong interest to understand the dynamical processes in the Local Interstellar
Cloud (LIC). They provide the main reservoir and transport mechanism of heavy 
elements in the interstellar medium \citep{li1997}. The dynamics of the grains
is crucial for understanding of nucleation, growth and collisional destruction
processes \citep{draine2003}. These processes strongly depend on the relative
velocities of the grains. The most important phenomena responsible for the spread of
velocities in the LIC are gas drag, interaction with the local 
interstellar magnetic field, radiation pressure and photoelectric
emission \citep{frisch1999a}. The relative strengths of the different forces
strongly depend on the size and the charge of the grains, together with local
conditions of the interstellar medium (ISM), like gas or magnetic field 
turbulences.


Observations of interstellar material
(ISM) towards nearby stars and inside of the solar system, combined
with radiative transfer models, give a self-consistent description of
the LIC 
\citep{frisch1998,frisch1999b,frisch2003a,slavin2006,slavin2007a,slavin2007b}.
The main characteristics of the LIC are: atomic neutral hydrogen
density $\mathrm{n(H^{o}) \sim 0.2\,cm^{-3}}$, electron and ion density 
$\mathrm{n(e^{-}) \sim 0.10\,cm^{-3}}$, temperature 
$\mathrm{\sim 6300\,K}$, and a relative
Sun-cloud velocity $\mathrm{\sim 26\,km \,s^{-1}}$.
The physical conditions in
the LIC are those of the intercloud medium --- warm, low density,
partially ionized gas. An enhancement of refractory elements (such
as Fe, Mg, Mn) in LIC gas, compared to cool interstellar clouds, points
to the destruction of interstellar dust grains by interstellar shocks
(velocity 100 to 200 km s$^{-1}$) \citep{frisch1999a}. 

The ISM within 10 pc of the Sun is highly inhomogeneous. 
At least 5 distinct cloudlets are found within 5 pc of the Sun, with
differing compositions and physical properties. Temperatures range
from 5,400 K (towards $\alpha$ Cen) to 10,000 K (Blue Cloud towards
$\epsilon$ CMa) and total densities from $\mathrm{ >0.04\,cm^{-3}}$ 
(Blue Cloud
towards $\epsilon$ CMa) to possibly $\mathrm{> 5\,cm^{-3}}$ 
\citep[G-cloud
towards $\alpha$ Cen,][]{frisch2003b,gry2001}. The gas-phase
abundance of Fe, with respect to undepleted S, varies by $\sim$50\%
within 3 pc of the Sun, evidently due to grain destruction processes
\citep{frisch2003a}.

If the ISM is chemically homogeneous, elements absent from the gas
phase must be depleted onto dust grains. This argument can be used to
evaluate the gas-to-dust mass ratio \Rgd{} over the integrated LIC 
column, and \Rgd{} can be compared with that of other
nearby interstellar clouds. However, the required knowledge of the
total chemical composition of the ISM is an elusive quantity that has
not been reliably determined. A 40\%--50\% variation in \FeII/\SII{}
and \SiII/\SII{} for the two clouds towards $\epsilon$ CMa indicates
different grain histories for two similar clouds within 3 pc of
each other. If atoms not observed in the gas are concentrated in the
dust, \Rgd{} can be calculated from observations of interstellar
absorption lines towards nearby stars. When evidence for 60\%--70\%
subsolar abundances is included, \Rgd$\sim$600 integrated over the
diameter of the LIC \citep{frisch1999a,frisch2003a}. Gas-to-dust mass
ratios calculated from more recent models with improved solar
abundances are in the range \Rgd$\sim$140--490, again depending on solar
abundances \citep{slavin2007b}. Interestingly, \Rgd{} determined from
comparisons of the Ulysses in-situ measurements inside 
of the solar system, compared to gas densities from these models, 
yield \Rgd = 116--127 \citep{landgraf2000a,altobelli2004a}. It should be
emphasised that the
\Rgd{} obtained from the in-situ measurements is an upper limit, 
since the smallest interstellar dust grains (radii 0.1 \mum) are
prevented from entering the heliosphere. 

Overall, the in-situ value is a factor 
of $>$ 2 larger than the one derived from astronomical observations, 
indicating a relative concentration of interstellar dust in the ISM 
close to the Sun compared to the $\sim$0.5 pc LIC cloud length 
towards $\epsilon$ CMa. The gas-to-dust
mass ratio also varies by more than 30\% over the nearest 3 pc. If ISM
abundances are solar, the in-situ and astronomical methods of
determining \Rgd{} are -- generally -- in better agreement, but 
interstellar absorption line data towards weakly reddened stars remain 
unexplained. These differences are not yet understood. The chemical 
composition of
interstellar dust grains observed within the solar system thus
provides a window on the chemical composition and homogeneity of the
ISM.

The combination of absorption line data toward $\epsilon$ CMa and the
modelled photoionization also lead to the conclusion that the LIC
has a very interesting pattern of gas phase elemental abundances
\citep{slavin2007a}: C appears to be substantially  supersolar
while Fe, Mg and Si are subsolar. O and N are close to solar. 
This indicates that carbonaceous grains have been destroyed in the
LIC while silicate grains have survived. The extra C in the gas
has not been explained but may be evidence for a local
enhancement of carbonaceous dust followed by grain destruction in 
a shock. 

The masses of interstellar grains measured in-situ with the spacecraft 
detectors range 
from $\mathrm{10^{-18}\,kg}$ to above $\mathrm{10^{-13}\,kg}$. If we compare the
mass distribution of these interstellar impactors with
the dust mass distribution derived from astronomical observations, we
find that the in-situ measurements overlap only with the largest 
masses observed astronomically. This is further supported by the
radar measurements which revealed even bigger grains. These 
measurements imply that
the intrinsic size distribution of interstellar grains in the LIC
extends to sizes much larger than those grains which are 
detectable by astronomical 
observations \citep{frisch1999a,frisch2003a,landgraf2000a,gruen2000b}.

There are no direct observations of interstellar dust within 5 pc and
outside of the solar system. The observations of very weak starlight
polarization towards nearby stars ($\mathrm{<}$40 pc) may originate from
magnetically aligned dust grains close to the solar system. The observed
polarization strength is consistent with the average interstellar
density of $\mathrm{\sim 0.1 cm^{-3}}$ over tens of parsecs in the upwind
direction \citep{frisch1990}. The in-situ grains have a size
distribution consistent with these classical dust grains.

Interstellar gas and dust couple through collisional processes, and
through coupling of ions and charged grains to the interstellar
magnetic field. Over distance scales of 100--500 pc, gas-dust
coupling is demonstrated through the correlation of
starlight-reddening dust grains (measured as color excess E(B-V)) and
interstellar hydrogen (\NHI+2\NH2)
\citep{bohlin1978}.
For the multicloud structure observed within 5 pc, gas-dust coupling
is not proven. The collisional lifetimes for classical dust grains
(radii $\sim0.2\mu$m) in the LIC are $\sim0.3\times 10^6$ years, during
which time the LIC will move $\sim$5 pc through local space. The
gyroradius is $\sim$0.1 pc in a field of 3 $\mu$G
\citep{gruen2000b}. The result will be magnetically captured
dust grains that are collisionally destroyed over the lifetime of the
cloud. Gr\"un and Landgraf (2000) suggested that the small ``classical'' 
grains are
replenished by the collisional destruction of larger dust grains.
Alternatively, silicate grain destruction peaks near shock column
densities of $\mathrm{N(H)\sim 6 \times 10^{17}\,cm^{-2}}$ \citep{jones1994},
allowing the breakdown of gas-dust coupling locally over $\sim$1 pc
length scales. 

There are important consequences from the existence of the big particle population in 
the LIC. While particles observed by spacecraft couple to the interstellar medium 
on length scales of less than 1 pc via electromagnetic 
interactions, more massive grains couple to the gas over much longer scales 
of 100 to 1000~pc \citep{gruen2000b}. Therefore, big interstellar meteor particles 
travel unaffected over much longer distances and may come directly from their source 
region.

\section{Outlook}

Ulysses is presently in the
inner solar system where interstellar grains cannot reliably 
be separated from interplanetary impactors. After mid-2008, however, 
if the Ulysses spacecraft remains in good health, the dust instrument will 
monitor the interstellar grains again. 
The Ulysses mission is presently planned to be extended until at least 
early 2009 so that additional dust data will hopefully become available. A 
further mission extension until 2011 
is technically feasible and may provide interstellar dust data from 
the outer heliosphere again. With this latter extension the Ulysses 
measurements would cover an almost entire 22-year
solar cycle. It would make the Ulysses data a unique data set 
of dust measurements from interplanetary space for decades to come.
Together with detailed modelling of the grain interaction with the 
IMF during the highly disordered solar maximum conditions we will
hopefully be able to reveal the origin of the observed $\mathrm{30^{\circ}}$
shift. If the shift turned out 
to be intrinsic, being potentially connected with a secondary population of 
interstellar grains, it would put strong constraints on the small-scale 
structure of the LIC. 
This would also be highly relevant for the interpretation of results
from the Stardust mission which recently brought a sample of collected 
interstellar grains to Earth (A. Westphal 2006, priv. comm.),  and for future 
dust astronomy space missions aiming at the in-situ analysis and sample return of
interstellar dust \citep[CosmicDUNE, SARIM;][]{gruen2005,srama2008}.

\begin{acknowledgements}
We thank the Ulysses project at ESA and NASA/JPL for 
effective and successful mission operations. 
This work has been supported by the Deutsches Zentrum f\"ur 
Luft- und Raumfahrt e.V. (DLR) under grants 50 0N 9107 and 50 QJ 9503. 
Support by Max-Planck-Institut f\"ur Kernphysik and Max-Planck-Institut f\"ur 
Sonnensystemforschung is also gratefully acknowledged.
\end{acknowledgements}

\newcommand{\bibfont}{\footnotesize}
\bibliographystyle{SSRv}


\end{document}